\documentclass[iop]{emulateapj}
\slugcomment{Accepted by ApJ Letters, December 30, 2014} 
\usepackage{amsmath,hyperref,breakurl}

\shorttitle{Outbursting Class 0 Protostar}
\shortauthors{Safron et al.}

\begin{document}

\title{HOPS 383: An Outbursting Class 0 Protostar in Orion}
\author{
Emily J. Safron\altaffilmark{1},
William J. Fischer\altaffilmark{2},
S. Thomas Megeath\altaffilmark{1},
Elise Furlan\altaffilmark{3},
Amelia M. Stutz\altaffilmark{4},
Thomas Stanke\altaffilmark{5},\\
Nicolas Billot\altaffilmark{6},
Luisa M. Rebull\altaffilmark{3},
John J. Tobin\altaffilmark{7},
Babar Ali\altaffilmark{8},
Lori E. Allen\altaffilmark{9},
Joseph Booker\altaffilmark{1},\\
Dan M. Watson\altaffilmark{10}, and
T. L. Wilson\altaffilmark{11} 
}
\altaffiltext{1}{Ritter Astrophysical Observatory, Department of Physics and Astronomy, University of Toledo, Toledo, OH, USA; ejsafron@gmail.com}
\altaffiltext{2}{NASA Postdoctoral Program Fellow, Goddard Space Flight Center, Greenbelt, MD, USA; william.j.fischer@nasa.gov}
\altaffiltext{3}{Infrared Processing and Analysis Center, Caltech, Pasadena, CA, USA}
\altaffiltext{4}{Max-Planck-Institut f\"ur Astronomie, Heidelberg, Germany}
\altaffiltext{5}{European Southern Observatory, Garching bei M\"unchen, Germany}
\altaffiltext{6}{Instituto de Radio Astronom\'ia Milim\'etrica, Granada, Spain}
\altaffiltext{7}{Leiden Observatory, Leiden, Netherlands}
\altaffiltext{8}{Space Science Institute, Boulder, CO, USA}
\altaffiltext{9}{National Optical Astronomy Observatory, Tucson, AZ, USA}
\altaffiltext{10}{Department of Physics and Astronomy, University of Rochester, Rochester, NY, USA}
\altaffiltext{11}{Naval Research Laboratory, Washington, DC, USA}


\begin{abstract}
We report the dramatic mid-infrared brightening between 2004 and 2006 of HOPS 383, a deeply embedded protostar adjacent to NGC 1977 in Orion.  By 2008, the source became a factor of 35 brighter at 24 \micron\ with a brightness increase also apparent at 4.5 \micron.  The outburst is also detected in the submillimeter by comparing APEX/SABOCA to SCUBA data, and a scattered-light nebula appeared in NEWFIRM $K_s$ imaging.  The post-outburst spectral energy distribution indicates a Class 0 source with a dense envelope and a luminosity between 6 and 14 $L_\sun$.  Post-outburst time-series mid- and far-infrared photometry shows no long-term fading and variability at the 18\% level between 2009 and 2012.  HOPS 383 is the first outbursting Class 0 object discovered, pointing to the importance of episodic accretion at early stages in the star formation process.  Its dramatic rise and lack of fading over a six-year period hint that it may be similar to FU Ori outbursts, although the luminosity appears to be significantly smaller than the canonical luminosities of such objects.
\end{abstract}


\section{INTRODUCTION}

Early infrared studies of large samples of young stars revealed that protostars were under-luminous compared to predictions \citep{ken90}.  In order to form a solar-mass star in $10^5$ yr, the time-averaged mass accretion rate onto the star over this period needs to be $10^{-5}~M_\sun~{\rm yr}$, implying total luminosities in excess of 10 $L_\sun$.  Median protostellar luminosities, however, are of order 1 $L_\sun$ \citep{ken90,eva09,kry12}.  One means of resolving the discrepancy is {\em episodic accretion}, in which the luminosity of a forming star is usually $\sim1$ $L_\sun$ as observed, but a series of relatively brief, dramatic spikes in the accretion rate and luminosity over the star-formation period supply the requisite mass for a sun-like star \citep{har96}.

During the 20th century, several young stars were observed to undergo bursts consistent with the episodic accretion hypothesis, beginning with FU Ori in 1936 \citep{wac39}.  With extensive optical to infrared surveys of star-forming regions by, e.g., the Palomar Transient Factory and the {\em Spitzer Space Telescope}, and with careful work by amateur astronomers such as R.\ Persson (V733 Cep), J.\ McNeil (V1647 Ori), and A.\ Jones (EX Lup), many additional outbursts have been discovered over the past decade.  The recent review of \citet{aud14} catalogs 26 eruptive young stars.

While the first outbursts to be observed resembled Class II young stellar objects (YSOs), with luminous accretion disks but weak to absent circumstellar envelopes, further discoveries have extended the outburst phenomenon to the envelope-embedded Class I protostars.  Such protostellar outbursts include V346 Nor \citep{gra85}, OO Ser \citep{hod96,kos07}, V1647 Ori \citep{rei04,abr04,bri04}, V2775 Ori \citep{car11,fis12}, and V900 Mon \citep{rei12}.  By obtaining 70 \micron\ light curves for protostars in the Orion Nebula Cluster, \citet{bil12} also found evidence for variable accretion below the level associated with bursts, but still in excess of the 10\% level, in eight of 17 protostars.

The relative importance of episodic accretion to the star-formation process is still under debate \citep{bar09,hos11}.  Is the majority of the mass of a typical solar-mass star accreted in stochastic outbursts or in a smooth process of infall from a circumstellar envelope, through an accretion disk, and onto the star?  The likelihood of the former scenario increases as new outbursts are discovered at earlier stages of the star formation process.  Here we announce the discovery of an outburst in HOPS 383, a Class 0 protostar with the reddest spectral energy distribution (SED) yet observed for an outburst, implying that it may be the youngest known episodic accretor.

\section{OBSERVATIONS}

In Orion, we performed a search for variability consistent with episodic accretion by comparing photometry of protostars at 3.6, 4.5, and 24 \micron\ from the 2004--2005 {\em Spitzer} survey by \citet{meg12} to photometry at 3.4, 4.6, and 22 \micron\ from the 2010 survey of the {\em Wide-Field Infrared Survey Explorer} ({\em WISE}; \citealt{wri10}).  While the full results of this search will be presented in a subsequent publication, here we focus on the dramatic outburst of HOPS 383.

HOPS 383, at $\alpha=5^h35^m29^s.81$, $\delta=-4^\circ59'51''.1$ (J2000), was identified as a protostar by \citet{meg12} and is a source targeted by HOPS, the {\em Herschel} Orion Protostar Survey \citep[e.g.,][]{stu13,man13}.  We assume it is at a distance of 420 pc, the same as adopted by \citet{meg12} for the entire Orion complex.

\subsection{Pre-Outburst Observations}

On 2000 December 12, \citet{pet08} obtained a $K$-band image of HOPS 383 with SQIID, the Simultaneous Quad Infrared Imaging Device at the Kitt Peak 2.1 m telescope.  In 2004, HOPS 383 was detected at 4.5 \micron\ by the Infrared Array Camera (IRAC) and at 24 \micron\ by the Multiband Imaging Photometer (MIPS) aboard {\em Spitzer}.  The IRAC observations were obtained in two epochs, one on 2004 March 9 and the other on 2004 October 12, and the MIPS observation was on 2004 March 20.  An accounting of the {\em Spitzer} observations and subsequent analysis can be found in \citet{kry12} and \citet{meg12}.  Color corrections for the pre- and post-outbust IRAC fluxes, which were determined for a flat-spectrum source, were estimated to be minimal.  The pre- and post-outbust MIPS data were corrected by $+5\%$ relative to a Rayleigh-Jeans flux law due to the redness of the source.  We also present a 450~\micron\ image from SCUBA, the Submillimetre Common-User Bolometer Array, that was obtained in 1998 \citep{joh99}.

\subsection{Post-Outburst Observations}

A $K_s$-band image of HOPS 383 was obtained on 2009 November 25 with NEWFIRM, the NOAO Extremely Wide-Field Infrared Imager, at the Kitt Peak 4 m telescope.  A region including HOPS 383 was observed with the {\em Spitzer} InfraRed Spectrograph (IRS) in mapping mode on 2006 October 20.  We performed photometry of HOPS 383 on the peak-up images acquired at 15.8 and 22.3 \micron; details will be provided in our forthcoming publication about the larger variability search.  A second MIPS observation was acquired on 2008 April 19.  As part of the YSOVAR program \citep[Young Stellar Object VARiability;][]{mor11,reb14}, HOPS 383 was observed 81 times in the fall of 2009 and 11 times in the fall of 2010 with the IRAC 3.6 and 4.5 \micron\ channels.

{\em WISE} data were acquired in two visits on 2010 March 8--9 and 2010 September 15--16.  The first visit obtained photometry at 3.4, 4.6, and 22 \micron\ and upper limits at 12 \micron.  The second visit obtained photometry only at 3.4 and 4.6 \micron.  We applied color corrections to the photometry by fitting the model SED (\S3.2) within each bandpass with a Planck function (bands 1 and 2) or a power law (bands 3 and 4) and interpolating between cases listed in the {\em WISE} Explanatory Supplement.\footnote{See \url{http://wise2.ipac.caltech.edu/docs/release/allsky/expsup/sec4_4h.html}.} Including the additional correction recommended at 22 \micron, the corrections were $-18\%$, $-6.1\%$, $+1.3\%$, and $-12\%$ in bands 1 though 4.

With {\em Herschel} we observed HOPS 383 on 2010 September 10 and 28 in the 70~$\mu$m and 160 $\mu$m bands of the Photodetector Array Camera and Spectrometer (PACS).  These observations are Groups 19 and 135 in Table 2 of \citet{stu13}, who give additional information about the observations and describe the data processing and aperture photometry.  We also include 100 \micron\ photometry from the {\em Herschel} Gould Belt Survey \citep{and10} observation of 2010 October 8 as well as 18 {\em Herschel}/PACS 70 \micron\ data points obtained between 2011 February 25 and 2012 August 27 \citep{bil12}.

For submillimeter coverage, we report beam fluxes from our Atacama Pathfinder Experiment (APEX) survey of Orion.  We acquired a 350~\micron\ image on 2011 September 16 with SABOCA, the Submillimetre APEX Bolometer Camera, and an 870 \micron\ image on 2010 October 24 with LABOCA, the Large APEX Bolometer Camera.  Details of the APEX observations, data reduction, and photometry are reported in \citet{stu13}.

\section{RESULTS}

\subsection{Detection and Timing Constraints}

\begin{figure*}
\resizebox{\hsize}{!}{\includegraphics{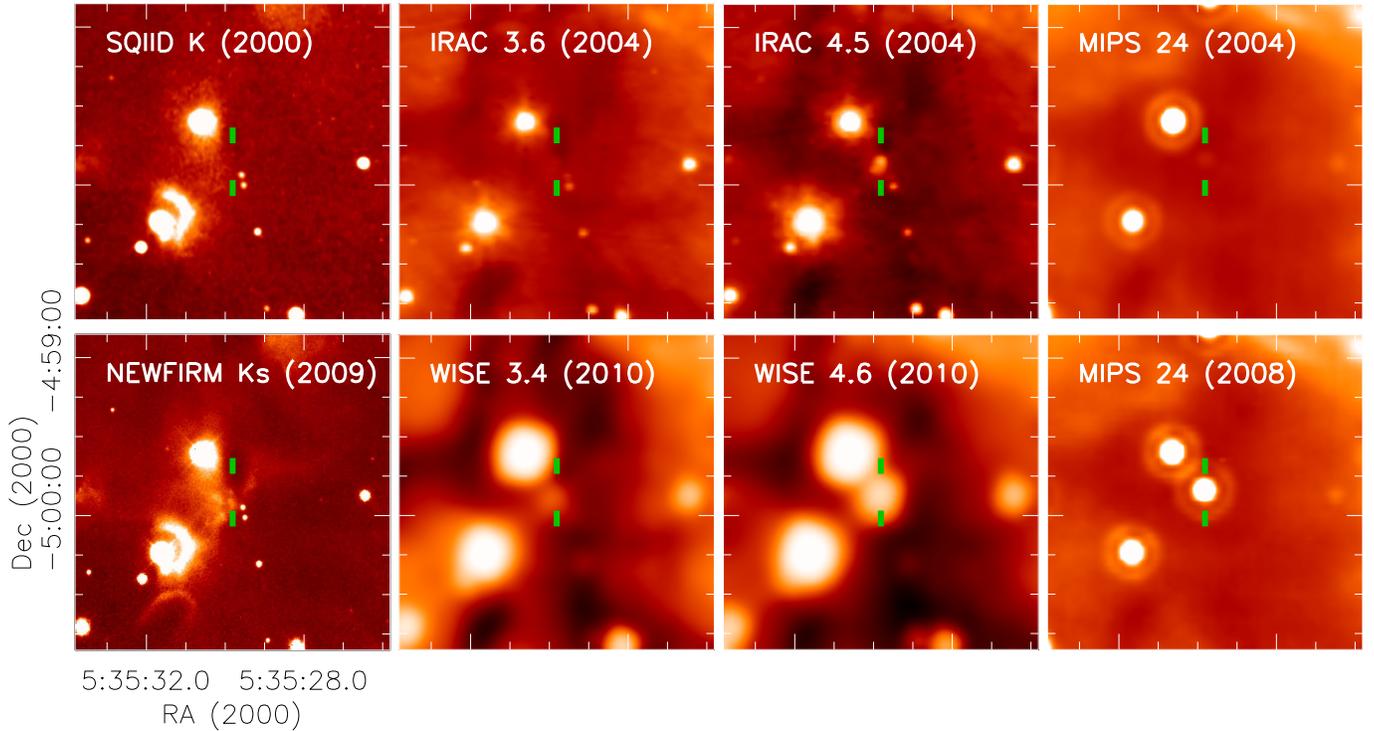}}
\figcaption{Near- and mid-IR images of HOPS 383 before and after its luminosity increase. {\em Top:} Pre-outburst imaging from SQIID and {\em Spitzer}.  {\em Bottom:} Post-outburst imaging from NEWFIRM, {\em WISE}, and {\em Spitzer}.  The position of HOPS 383 is marked in green.\label{f.images}}
\end{figure*}

Fluxes for HOPS 383 before and after the outburst appear in Table~\ref{t.phot}, while Figure~\ref{f.images} shows near- and mid-infrared images.  In the $K$ band, the point source is not detected at either epoch, but nebulosity appears post-outburst. We report fluxes inside an aperture of radius 4.8\arcsec, the largest radius that does not include significant emission from nearby stars, with subtraction of the median sky signal in an annulus extending from 52\arcsec\ to 60\arcsec\ that is free from bright point sources or extended emission.  These fluxes may include contamination from nebulosity in the region unrelated to HOPS 383, but this is likely to be constant across the two epochs, indicating an increase of 0.476 mJy at $K$ due to the outburst.

\begin{deluxetable}{ccccc}
\tablecaption{Photometry for HOPS 383\label{t.phot}}
\tablewidth{\hsize}
\tablehead{\colhead{$\lambda$} & \colhead{$F_\nu$} & \colhead{$\sigma F_\nu$} & \colhead{Instrument} & \colhead{Date} \\ \colhead{(\micron)} & \colhead{(mJy)} & \colhead{(mJy)} & \colhead{} & \colhead{}}
\startdata
2.2  & 0.374 & 0.0551 & SQIID & 2000 Dec 12\tablenotemark{1}\\
2.2 & 0.850 & 0.0272 & NEWFIRM & 2009 Nov 25\tablenotemark{1}\\
3.4  & 0.585 & 0.0969 & WISE & 2010 Mar 8--Sep 16\\
3.6  & 1.54 & 0.0770 & IRAC & 2009 Oct 23\tablenotemark{2} \\
4.5  & 1.30 & 0.0770 & IRAC & 2004 Mar 9--Oct 12 \\
4.5  & 6.34 & 0.317 & IRAC & 2009 Oct 23\tablenotemark{2} \\
4.6  & 10.5 & 0.358 & WISE & 2010 Mar 8--Sep 16 \\
12  & $<16.5$ & \nodata & WISE & 2010 Mar 8--9 \\
16 & 3.50 & 0.450 & IRS & 2006 Oct 20 \\
22 & 80.5 & 3.32 & IRS & 2006 Oct 20 \\
22 & 198 & 16.8 & WISE & 2010 Mar 8--9 \\
24  & 5.34 & 0.282 & MIPS & 2004 Mar 20 \\
24  & 187 & 9.87 & MIPS & 2008 Apr 19 \\
70   & 13100 & 150 & PACS & 2010 Sep 10\tablenotemark{2} \\
100 & 25400 & 1360 & PACS & 2010 Oct 8 \\
160 & 35100 & 2010 & PACS & 2010 Sep 10--28 \\
350 & 6620 & 2650 & SABOCA & 2011 Sep 16 \\
870 & 996 & 199 & LABOCA & 2010 Oct 24
\enddata
\tablenotetext{1}{Flux in an aperture of radius 4.8\arcsec\ with a sky-subtraction annulus from 52\arcsec\ to 60\arcsec.  It likely contains a contribution from nebulosity not associated with HOPS 383; however, this should be constant between the two epochs.}
\tablenotetext{2}{Representative of a time series.}
\end{deluxetable}

The best demonstration of the change in the point source appears in the MIPS images, where the source became a factor of 35 brighter at 24 \micron\ between 2004 and 2008.  The 2006 IRS measurement at 22 \micron\ is much closer to the 2008 MIPS flux density than the 2004 MIPS flux density; therefore, we conclude the outburst began between 2004 October 12 and 2006 October 20, with a subsequent rise in luminosity up to 2008.

\begin{figure}
\resizebox{\hsize}{!}{\includegraphics[viewport=0 20 400 250]{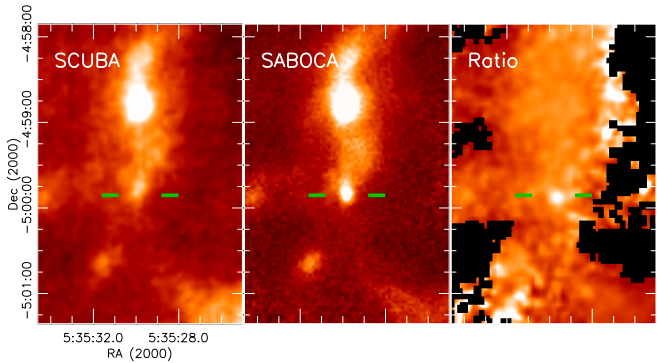}}
\figcaption{Submillimeter images. A SCUBA image at 450 \micron\ obtained in 1998 appears on the left, a SABOCA image at 350 \micron\ obtained in 2011 appears in the center, and the ratio of the post-outburst to the pre-outburst image appears on the right.  The position of HOPS 383 is marked in green.\label{f.smm}}
\end{figure}

Figure~\ref{f.smm} shows a SCUBA 450 \micron\ image from 1998, a SABOCA 350 \micron\ image from 2011, and the ratio of SABOCA to SCUBA.  They show the appearance by 2011 of a bright source at the position of HOPS 383. While the extended emission has a ratio of $\sim2.1$ due to the difference in wavelength, the source has a ratio of $\sim4$, implying an increase in brightness of $\ge2$.

\begin{figure}
\resizebox{\hsize}{!}{\includegraphics{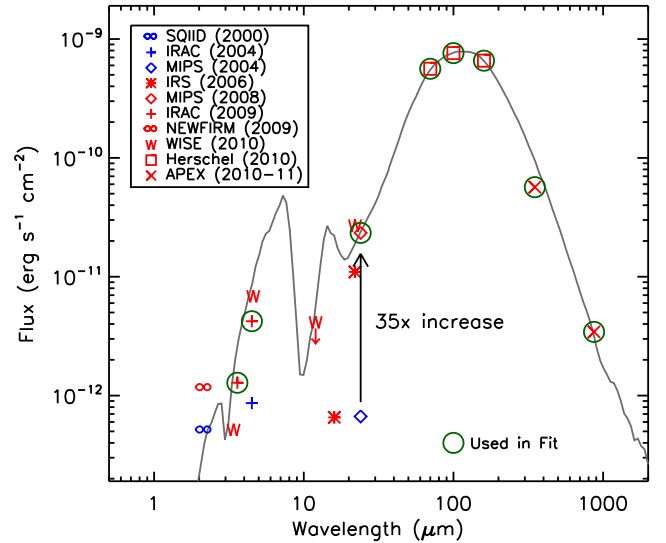}}
\figcaption{Pre- and post-outburst photometry for HOPS 383, with pre-outburst data in blue and post-outburst data in red.  The best-fit SED from the HOPS grid of radiative transfer models is shown with a gray curve.  The {\em WISE} point with an arrow below it is an upper limit.  For data products where time series exist, a representative flux is shown.  Green circles mark the data used in modeling.\label{f.sed}}
\end{figure}

The SED of HOPS 383 appears in Figure~\ref{f.sed}.  Due to the paucity of pre-outburst data, we focus on the post-outburst SED.  For data products where time series exist, the variability within the time series is small compared to the range of the whole SED, so we choose a representative observation for the SED; the time series are discussed in Section 3.3.  

\subsection{Properties of the Post-Outburst Source}

The bolometric luminosity $L_{\rm bol}$ of the post-outburst SED is 7.5 $L_\sun$.  \citet{dun08} showed that, for protostars, the correlation between flux density at a given wavelength and luminosity tightens with increasing wavelength between 3 and 70 \micron.  Lacking pre-outburst 70 \micron\ data, we estimate that the change in luminosity was equal to the factor of 35 increase in the 24 \micron\ flux density, giving a pre-outburst luminosity of 0.2 $L_\sun$.

A common diagnostic of evolutionary state is the ratio of submillimeter luminosity $L_{\rm smm}$ to bolometric luminosity.  The Class 0 objects have $L_{\rm smm}/L_{\rm bol}>0.5\%$ \citep{and93}, and HOPS 383 has a ratio of 1.4\%, confirming its deeply embedded status and making it the only known Class 0 accretion outburst.  This ratio is toward the low end of the range observed for the extreme Class 0 objects known as PACS Bright Red Sources \citep{stu13}.

Another diagnostic is the bolometric temperature, the temperature of a blackbody with the same mean frequency as that of the protostellar SED \citep{mye93}.  The Class 0 protostars have $T_{\rm bol}<70~{\rm K}$ \citep{che95}, and these correspond roughly to protostars in which the majority of the mass is in the infalling envelope, not yet in the star \citep{dun14}.  The SED of HOPS 383 has $T_{\rm bol}=43~{\rm K}$, again consistent with Class 0.

Additional evidence for a deeply embedded object comes from fitting a modified blackbody to the SED, where the method of \citet{stu13} gives a peak wavelength of 106 \micron\ (larger peak wavelengths imply denser envelopes) and a lower limit to the envelope mass of 0.2 $M_\sun$.  Estimating the mass from the 870 \micron\ APEX flux alone, assuming a temperature of 18 K (Stutz et al., in preparation) and OH5 opacities from \citet{oss94}, gives a mass of 0.7 $M_\sun$ in the LABOCA beam, which has a half-width at half maximum corresponding to a radius of 4000 AU.  These mass estimates point to a protostar in the earlier phases of infall.

To estimate the luminosity, inclination angle, and cavity opening angle of HOPS 383, we modeled the {\em Spitzer}, {\em Herschel}, and APEX photometry of the source with the radiative transfer code of \citet{whi03}.  The HOPS team created a grid of 3040 model SEDs with parameters appropriate for protostars, first described by \citet{ali10}, and updated by E.\ Furlan et al.\ (in preparation).  We found the best fit by minimizing $R$, which measures the logarithmic deviation of the models from the observations in units of the fractional uncertainty \citep{fis12}.

The best-fit model ($R=1.7$) has a total luminosity of 8.7 $L_\sun$, an inclination of 41$^\circ$ from pole-on, and a cavity opening angle of 25$^\circ$.  Other models that provide satisfying fits to the data ($R<2.1$) have total luminosities ranging from 6 to 14 $L_\sun$, inclinations ranging from 41$^\circ$ to 63$^\circ$ from pole-on, and cavity opening angles from 15$^\circ$ to 35$^\circ$.  (The total luminosity can differ from the observed luminosity due to the non-isotropic radiation field.)  The scattered-light cone extending to the northwest and bright features to the southeast in the NEWFIRM image of Figure~\ref{f.images} appear to be from radiation escaping the outflow cavity and imply an intermediate inclination angle, consistent with the SED.

\subsection{Post-Outburst Variability}

\begin{figure}
\resizebox{\hsize}{!}{\includegraphics{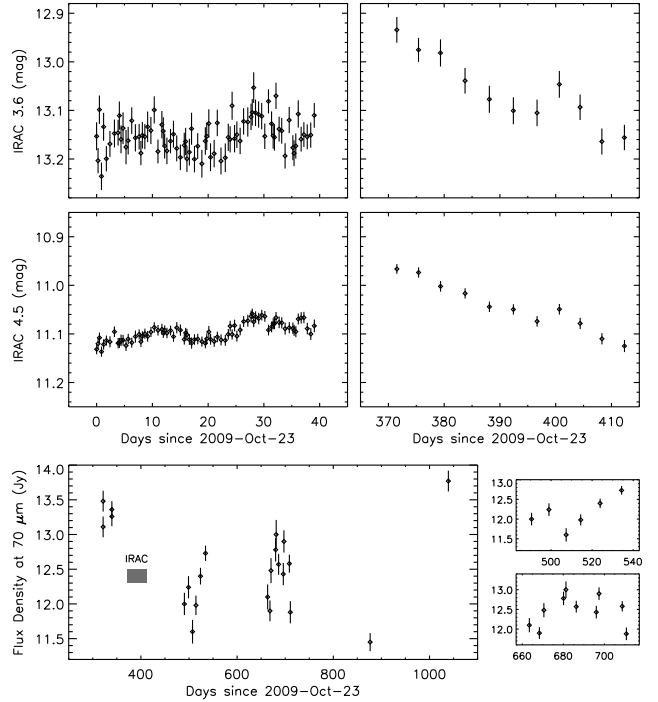}}
\figcaption{Time-series photometry. {\em Top two rows}: Magnitude versus time for the YSOVAR 2009 ({\em left}) and 2010 ({\em right}) campaigns.  {\em Bottom row}:  Flux density versus time for the {\em Herschel}/PACS 70 \micron\ data. The gray rectangle marks the epoch of the 2010 YSOVAR campaign, and the two small panels zoom in on the two 2011 PACS epochs.\label{f.var}}
\end{figure}

Our best sampling of the variability of HOPS 383 comes from the {\em Spitzer} YSOVAR data at 3.6 and 4.5 \micron\ and the {\em Herschel}/PACS 70 \micron\ data.  Figure~\ref{f.var} shows these light curves.  After the 2 mag jump at 4.5 \micron\ between 2004 and 2009 (not shown), the 2009 season of YSOVAR data yielded remarkably constant photometry, failing the variability tests laid out in \citet{reb14}.  The slopes of the best-fit lines to the light curves indicate brightenings of only 0.07 mag at 3.6 \micron\ and 0.08 mag at 4.5 \micron, and the $[3.6]-[4.5]$ color reddened by only 0.01 mag. 

There were two day-long series of {\em WISE} observations at 3.4 and 4.6 \micron\ between the 2009 and 2010 YSOVAR campaigns, on 2010 March 8--9 and 2010 September 15--16.  No significant variability from the {\em WISE} fluxes shown in the SED (Fig.~\ref{f.sed}) was detected.  The second (2010) YSOVAR season showed a slight fading of the source, with a dimming of 0.20 mag at 3.6 \micron\ and 0.15 mag at 4.5 \micron.  Again, the color was more consistent than the magnitudes, with a reddening of 0.05 mag over the window.

The {\em Herschel}/PACS 70 \micron\ data show variability at the 18\% level.  The first significant gap in the PACS observations coincides with the 2010 YSOVAR window; both light curves suggest a decline over this period.  Between the first 70 \micron\ observation on 2010 September 10 and the second-to-last one on 2012 March 18, the flux density declined by 13\% before recovering on 2012 August 27 to its brightest yet.

\section{DISCUSSION}

We have presented multi-epoch infrared data from 2.2 to 870 \micron\ to show an outburst of the Class 0 protostar HOPS 383, adjacent to the NGC 1977 nebula in Orion, between 2004 and 2006.  By 2008, the source became 35 times brighter at 24 \micron\ than it was in 2004.  The post-outburst luminosity, which is uncertain due to modeling degeneracies, is most likely between 6 and 14 $L_\sun$.  Subsequent monitoring of the source at 3.6, 4.5, and 70 \micron\ finds variability but no evidence for a significant decline in the source luminosity between 2009 and 2012.

One possible mechanism for the observed rise in the 24 \micron\ flux density would be the removal from the line of sight of a large amount of extinguishing material.  We are able to rule this out on two grounds.  First, using the opacity law from \citet{orm11} plotted in Figure~3 of \citet{fis14} and a gas-to-dust ratio of 100, a factor of 35 increase in the 24 \micron\ flux density would correspond to a reduction in the column density of $1.3\times10^{23}$ cm$^{-2}$, which would correspond to an implausible reduction in $A_V$ of 70.  Second, the source became brighter at submillimeter wavelengths (Fig.~\ref{f.smm}), which cannot be the result of a drop in extinction and is most likely due to an increase in the envelope temperature due to increased heating \citep{joh13}.  Ruling out a change in extinction implies that the observed flux increase was due to a genuine increase in the source luminosity.

YSOs undergoing such luminosity outbursts are generally classified as EX Lup objects, with repeated short-term flux increases of about a magnitude that persist for approximately one year, or as FU Ori objects, with a single burst of several magnitudes and the persistence of an elevated state for decades \citep{rei10}.  In both cases, an increase in the accretion rate onto the star is thought to be responsible for the luminosity increase. Outbursts have also been observed with light curves that do not fit cleanly into either category; e.g., V1647 Ori \citep{asp11} and V2492 Cyg \citep{cov11,hil13}, and final confirmation of the type of outburst requires optical or near-infrared spectra to diagnose conditions in the region where disk material is accreting onto the star \citep{con10}.  Due to the lack of such spectra for a deeply embedded source, we are limited in our ability to conclusively identify the class of outburst or the precise mechanism for the luminosity increase in HOPS 383, but the post-outburst light curves indicate the persistence of elevated luminosity from the first evidence of brightening in 2006 to the most recent {\em Herschel}/PACS imaging in 2012, inconsistent with the short-term EX Lup events.

HOPS 383 is unambiguously a Class 0 YSO based on the arguments presented in Section 3.2.  While outbursting YSOs are known to have envelopes in some cases \citep{qua07,gre13}, none has yet been detected that is as deeply embedded as HOPS 383.  Detections of disks in such deeply embedded protostars remain rare \citep{tob12,tob13}, but there is indirect evidence from the ubiquity of outflows and jets that there are disks around nearly all protostars \citep[e.g.,][]{fra14}.  Since proposed mechanisms for accretion outbursts are grounded in disk instabilities \citep[e.g.,][]{vor10,zhu10}, the outburst of HOPS 383 is another form of indirect evidence for a disk in a Class 0 source.  The discovery of this outburst demonstrates that episodic accretion can occur very early in the star formation process, and we encourage follow-up observations to begin to understand the physics at work.

\acknowledgments

Support for this work was provided by the National Aeronautics and Space Administration (NASA) through awards issued by the Jet Propulsion Laboratory, California Institute of Technology (JPL/Caltech).  We include data from {\em Herschel}, a European Space Agency space observatory with science instruments provided by European-led consortia and with important participation from NASA. We use data from the {\em Spitzer Space Telescope} and the Infrared Processing and Analysis Center Infrared Science Archive, which are operated by JPL/Caltech under a contract with NASA. We also include data from the Atacama Pathfinder Experiment, a collaboration between the Max-Planck-Institut f\"ur Radioastronomie, the European Southern Observatory, and the Onsala Space Observatory. This paper makes use of data products from the {\em Wide-field Infrared Survey Explorer}, which is a joint project of the University of California, Los Angeles, and JPL/Caltech, funded by NASA. This paper uses observations taken at Kitt Peak National Observatory, National Optical Astronomy Observatory, which is operated by the Association of Universities for Research in Astronomy under cooperative agreement with the National Science Foundation. The work of W.~F.\ was supported by an appointment to the NASA Postdoctoral Program at Goddard Space Flight Center, administered by Oak Ridge Associated Universities through a contract with NASA. The work of A.~S.\ was supported by the Deutsche Forschungsgemeinschaft priority program 1573 (``Physics of the Interstellar Medium'').

{\it Facilities:} \facility{APEX}, \facility{Herschel}, \facility{JCMT}, \facility{KPNO:2.1m}, \facility{Mayall}, \facility{Spitzer}, \facility{WISE}

\end{document}